\documentclass[sigconf]{acmart}

\usepackage{amsmath}
\usepackage{graphicx}
\usepackage{multirow}
\usepackage{xcolor}
\usepackage{algorithm}
\usepackage[noend]{algpseudocode}
\usepackage{comment}
\usepackage{placeins}

\AtBeginDocument{%
  \providecommand\BibTeX{{%
    \normalfont B\kern-0.5em{\scshape i\kern-0.25em b}\kern-0.8em\TeX}}}

\setcopyright{none}

\renewcommand\footnotetextcopyrightpermission[1]{} 
\begin{document}

\def \shai #1{\color{red}#1\color{black}}
\def \efrat #1{\color{blue}#1\color{black}}

\title{RadArnomaly: Protecting Radar Systems from Data Manipulation Attacks}


\author{Shai Cohen, Efrat Levy, \\ Yuval Elovici, Asaf Shabtai}

\affiliation{Department of Software and Information Systems Engineering, Ben-Gurion University of the Negev
\country{Israel}}

\author{Tair Cohen, Avi Shaked}

\affiliation{IAI ELTA Systems Cyber DIvision
\country{Israel}}

\renewcommand{\shortauthors}{Shai Cohen, et al.}


\begin{abstract}
Radar systems are mainly used for tracking aircraft, missiles, satellites, and watercraft.
In many cases, information regarding the objects detected by the radar system is sent to, and used by, a peripheral consuming system, such as a missile system or a graphical user interface used by an operator. Those systems process the data stream and make real-time, operational decisions based on the data received.
Given this, the reliability and availability of information provided by radar systems has grown in importance.
Although the field of cyber security has been continuously evolving, no prior research has focused on anomaly detection in radar systems.
In this paper, we present a deep learning-based method for detecting anomalies in radar system data streams.
We propose a novel technique which learns the correlation between numerical features and an embedding representation of categorical features in an unsupervised manner.
The proposed technique, which allows the detection of malicious manipulation of critical fields in the data stream, is complemented by a timing-interval anomaly detection mechanism proposed for the detection of message dropping attempts.
Real radar system data is used to evaluate the proposed method. Our experiments demonstrate the method's high detection accuracy on a variety of data stream manipulation attacks (average detection rate of 88\% with 1.59\% false alarms) and message dropping attacks (average detection rate of 92\% with 2.2\% false alarms).
\end{abstract}



\maketitle
\pagestyle{plain}

\section{\label{sec:intro}Introduction}

Radar systems use electromagnetic radiation to detect objects within a defined scanned area~\cite{maini2018handbook}; they can also be used to classify the detected objects.
Radar systems are mainly integrated in air and terrestrial traffic control systems, autonomous vehicles, air defense systems, anti-missile systems, aircraft anti-collision systems, and ocean surveillance systems.

Typically, radar system architecture consists of the following basic components: (1) an antenna - responsible for transmitting/receiving electromagnetic waves to/from the scanned area, and (2) a radar controller - responsible for analyzing the waves received from the antenna in order to determine the properties of the object. 

Usually, the radar controller is connected to a centralized switch which controls the routing of the radar system’s messages. 
In many cases, the switch can be used to communicate with external entities, i.e., other systems.
Often, information regarding the objects detected by the radar system is sent to, and used by, a peripheral consuming system, such as a missile system or a graphical user interface used by an operator~\cite{abdallah2017study}. 
Those systems process the data stream and make real-time, operational decisions based upon the data received. 

In recent years, as technology has evolved, the use of radar systems has increased, along with reliance on their correct and reliable operation. 
Radar systems often include an extended set of components, such as communication systems or SCADA systems, which are vulnerable to cyber and physical attacks.
These components can be exploited by attackers in order to compromise the radar system~\cite{cohen2019security}. 
In addition, in many cases, radar systems are integrated within systems that are vulnerable to cyber attacks, such as autonomous vessels~\cite{bolbot2020novel} and smart vehicles~\cite{kumar2018brief}. 
These vulnerabilities could be used by an attacker as a backdoor for an attack on the radar system. 

Despite the many potential attack vectors, there has been no research performed that has suggested or evaluated a method for detecting cyber attacks on radar systems in real time. In this study, we address this gap. 
We propose a deep learning-based method for detecting anomalies in critical parts of the data stream generated by the radar controller and consider attack scenarios in which the attacker manipulates the data being transferred by the radar system to peripheral consuming systems. 
Such manipulation can occur in motion or in use. 
The proposed method simply requires the ability to monitor the messages sent by the radar controller, and thus it can be easily and safely integrated into existing systems without the need to change the system.
Our proposed method is designed to deal with the particular data comprising the radar data stream; unlike many other domains, radar data streams are made up of sequential and heterogeneous data. 

In order to evaluate the proposed method, we used data collected from four real radar systems. 
We collected legitimate data streams into which we injected a variety of artificial attacks, including manipulations with high impact on the system (integrity violation and denial of service), to assess our method's performance. 
The results of our evaluation show that our proposed method can detect 90\% of message dropping attacks, with a false positive rate of 2\%, and from 76\% to 96\% of feature manipulation attacks, with a false positive rate of 2\%.
These results are obtained in a cross-session experiment in which the model is trained on data collected from three radar systems and tested on data collected from another radar system, thus demonstrating the ability to migrate a pretrained model to new radar systems without retraining.

We summarize the main contributions of this study as follows:
\begin{itemize}
    \item We present a novel deep learning-based method, which consists of two modules, to detect anomalies in critical parts of the heterogeneous data stream generated by radar systems.
    \item The proposed model represents both numerical and categorical features in an architecture that can be utilized in other similar domains. 
    \item The proposed method can be integrated into existing radar systems, without changing the radar system's existing components and without retraining the model.
    \item The structure of the proposed method enables it to identify the anomaly type detected, determine whether a legitimate message is missing from a sequence of messages, and understand whether a feature of an existing message has been manipulated.
    \item The effectiveness of the proposed method is demonstrated in experiments using a dataset collected from real operational radar systems and simulated attacks.
\end{itemize}

\section{\label{sec:setup}Background on Radar Systems}

A radar system uses radio waves to determine both the location of an object relative to the system and the distance between the object and the system.  
It operates by transmitting electromagnetic signals toward a certain direction and monitoring the signals that are reflected back from the objects. 
The reflected signals are then sent to a signal processor which analyzes the signals, aiming to extract the properties of the objects~\cite{abdallah2017study}. 

\begin{figure}[h]
\centering
\includegraphics[width=0.43\textwidth]{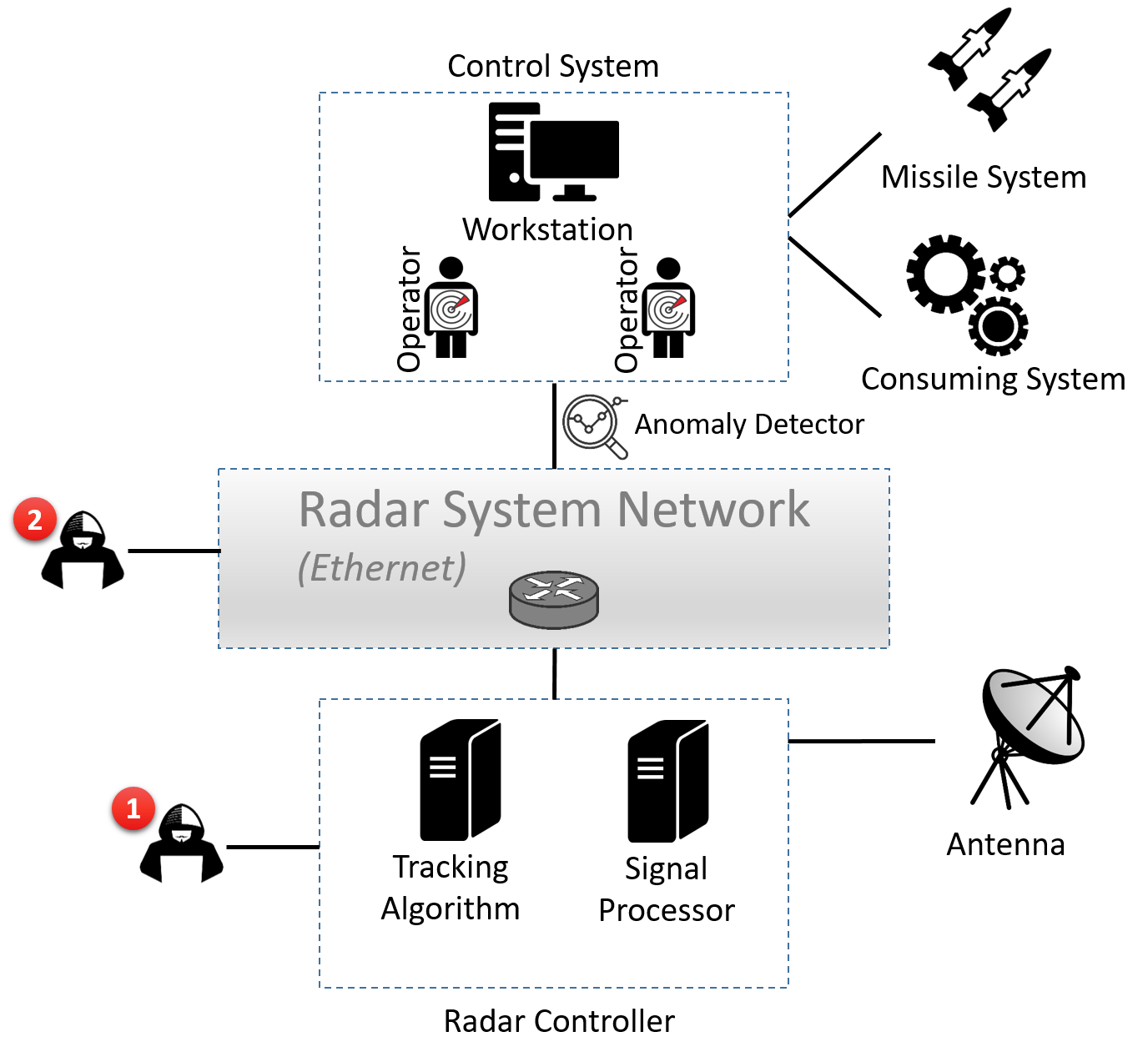}
\caption{An example of a typical radar system architecture. 
The red circles indicate our assumed adversary locations. 
Our proposed anomaly detection method is located at the link between the radar system and the consuming systems.}
\label{fig:system_description}
\end{figure}

\subsection{System Components}
A typical radar system architecture includes the following components (presented in Figure~\ref{fig:system_description}): \\
    \noindent \textbf{Antenna.} The antenna transforms the electric current into electromagnetic waves and vice versa (transmission and reception, respectively).
    Usually, one bidirectional antenna is used, however radar systems with two separate antennas (a transmitting antenna and a receiving antenna) also exist. \\
    \noindent \textbf{Radar Controller.} 
    The radar controller consists of two main components: 
    (1) a signal processor, which receives the signals from the antenna (usually via an optical link) and analyzes the signal using a signal processing algorithm aimed at identifying potential relevant objects, and
    (2) a tracking algorithm, which analyzes the signal processor's output, with the aim of classifying objects and tracking movement. \\
    \noindent \textbf{Control System.} The control system is responsible for analyzing radar yields and activating the different systems connected to the radar system, for example, activating a weapon system to neutralize a detected threat. \\
    \noindent \textbf{Radar System Network.} The radar system network is used as a communication channel between the radar controller and the control system. 

\subsection{\label{subsec:message_description}Data Stream Description}

The radar system (radar controller) continuously sends a stream of messages to peripheral consuming systems.
The consuming systems process the data stream and make real-time, operational decisions based on the data received.
Each message contains a \textit{Plot} record. This record describes a detected object at a given time.

\begin{figure}[h]
\centering
\includegraphics[width=0.34\textwidth]{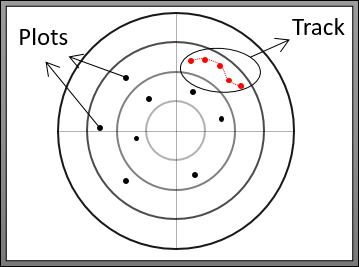}
\caption{Objects identified at a specific timestamp are represented by points. 
A sequence of points related to the same detected object are represented by a path of interconnected points (i.e., \textbf{\emph{Track}}).}
\label{fig:plots_Tracks}
\end{figure}

Since the radar system continuously scans the defined area, several $Plots$ may relate to the same object, describing its properties at different points in time. 
A sequence of $Plots$ related to the same object is defined as a $Track$. 
Each $Plot$ may be correlated with other $Plot$s (e.g., when they are part of an identified $Track$).
Each $Track$ is associated with an identification number (referred to as the TrackID). 
Thus, the TrackID is a part of each individual $Plot$ record.
Figure~\ref{fig:plots_Tracks} illustrates a set of $Plots$ (the individual points) and a $Track$ (the sequence of connected plots).

Each $Plot$ record contains the following attributes: 
\begin{itemize}
    \item \textbf{Metadata.} Contains the source ID (unique identifier of the sending radar system), message ID, message length, etc. 
    
    \item \textbf{Identified object's properties.} Categorical and numerical attributes that describe the identified object, such as the detected object's location and speed, object type (e.g., airplane, bird), etc. 

    \item \textbf{Sequence-related properties.} 
    Describe different properties of the $Plot$ record as part of a sequence of plots related to the same object.
    One property is the identification number of the $Track$ that a $Plot$ belongs to.
    Another property is the timestamp indicating when a $Plot$ has been updated on the system; this property describes the delay between the previous $Plot$ and the current $Plot$.
\end{itemize}

\section{\label{sec:threat_model}Assumed Threat Model}

Although radar systems are vulnerable to a variety of threats~\cite{cohen2019security}, most prior research performed on radar systems has focused on aspects of functionality~\cite{barton2004radar,barton1988modern}. 
In this study, we consider attack scenarios in which an attacker manipulates the data being transferred by the radar system to peripheral consuming systems. 
Such manipulation may occur in motion or in use. 
Specifically, as illustrated in Figure~\ref{fig:system_description}, the adversary locations (points of manipulation) considered in this study are the radar controller and the radar system network. 
Those can be compromised with physical or remote access, or through a supply chain attack (i.e., before the radar system's deployment). 

Once the radar controller has been compromised by the attacker (whether remotely or through a supply chain attack), he/she can execute malicious commands that manipulate the data while it is being processed, prior to its transmission from the radar controller. 
With physical access to the radar system network, the attacker can replace a networking device (e.g., switch) with a malicious networking device that manipulates the data in motion. The effects of such tampering attempts are expected to be covert and, more specifically, to be ``under the radar" of radar operators (i.e., radar operators should not be able to identify the tampering).

Based on the specified threat model, we suggest an anomaly detection method that (1) detects abnormal data behavior indicative of such malicious activity, and (2) can be implemented as a detector at the link between the radar system network and the control system, to provide anomaly-related insights to radar operators and/or systems that consume the radar data.

\section{\label{sec:related}Related Work}

Most studies that used machine learning techniques to protect radar systems from attacks focused on protecting the systems from jamming attacks. 
Some studies proposed jamming classification methods~\cite{hong2018classification,mendoza2017classification,soto2017optimization,wu2017jamming}, while others proposed jamming mitigation strategies~\cite{kang2018reinforcement,liu2020decentralized}.
In addition, a comprehensive survey discussing all of these methods in detail was published~\cite{lang2020comprehensive}. 
In other related research, machine learning techniques were used to protect the systems generally integrated within radar systems, like the ADS-B~\cite{ying2019detecting,habler2018using} and AIS systems~\cite{obradovic2014machine}.
To the best of our knowledge, no study has proposed a method for detecting anomalies in radar system data streams. 

In recent years, deep learning models have increasingly been used to analyze network traffic and detect anomalies. 
Most of the methods proposed are based on autoencoders and their variants~\cite{chalapathy2019deep}. 
An autoencoder (AE) is an unsupervised algorithm that compresses input into a lower dimensionality and then decompresses the
input into its original dimensionality; thus, normal instances are decompresses properly, while the abnormal instances are not. 
In this way, anomalous input can be identified.
Some examples of methods based on AEs include N-BIoT~\cite{meidan2018n}, which uses an AE to detect botnet behavior in the network traffic of IoT devices, and Kitsune~\cite{mirsky2018kitsune}, which utilizes a smart feature mapper and an ensemble of AEs to detect anomalous behavior in network traffic. 

LSTM-based networks are also widely used to detect anomalies in network traffic; these deep learning networks are very well suited for sequential data. 
In this type of network, instead of processing the whole sequence together, the network processes each object in the sequence separately, in chronological order. 
So in each step, the network's inputs are: (1) the current object in the sequence, along with (2) the output of the network from the previous step. 
An example of using an LSTM-based network was presented by Bontemps et al.~\cite{bontemps2016collective} who predicted anomalies in the network data stream by using the sample, as well as the context of the sample. 

Previous studies that used deep learning models to detect anomalies in network traffic relied strictly on numerical features. However, in this study, we take into consideration the fact that the data stream created by the radar system is heterogeneous, i.e., it contains both numerical and categorical features. Since manipulating both types of features may significantly violate the integrity and availability of radar systems, there is a need for a deep learning method that can detect anomalies in heterogeneous data.
Our proposed method is designed to detect attack scenarios that are valuable specially for radar systems. Thus, our proposed method focuses on real threats while reducing false alarms.


\section{\label{sec:method:high_level}High-Level Description of the Proposed Method} 

We propose an unsupervised anomaly detection method which is based on continuous monitoring of the messages transmitted from the radar to the peripheral consuming systems. 
Our proposed method utilizes state-of-the-art deep learning modules to detect possible malicious data manipulation. 

\subsection{Description of the Proposed Method}
\begin{figure}[h]
\centering
\includegraphics[width=0.48\textwidth]{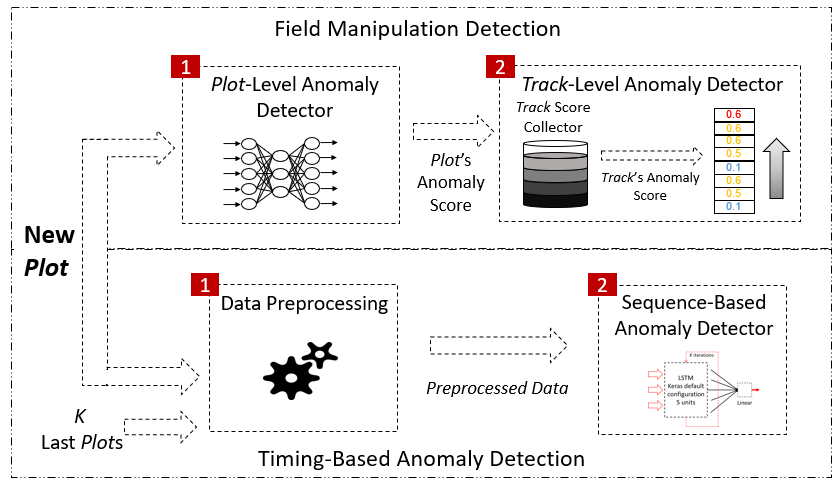}
\caption{High-level architecture of the proposed anomaly detection method.}
\label{fig:HLA}
\end{figure}

The proposed anomaly detection method (see Figure~\ref{fig:HLA}) consists of two main modules: (1) a Field Manipulation Detection module, and (2) a Timing-Based Anomaly Detection module. 
Each module outputs an anomaly score, and an alert is generated if the score exceeds a predefined threshold. 

\textbf{Field Manipulation Detection.} For each $Track$, the Field Manipulation Detection module receives the categorical and numerical features of a $Plot$. First, it determines if the relation between the features' values is anomalous. If so, an anomaly score for the $Plot$ is generated. 
An anomalous relation between the values of a $Plot$'s features indicates that a malicious change has been made, aimed at violating the radar system's integrity.
Such manipulations may cause the consuming systems or radar operators to incorrectly characterize and handle the detected objects.

Finally, to increase the sensitivity of the detection method to malicious manipulation attacks, an alert is generated if the scores' average exceeds a certain threshold.
The latter step aims to detect anomalies that are harder to detect by analyzing a single $Plot$.

\textbf{Timing-Based Anomaly Detection.} The Timing-Based Anomaly Detection module receives the last $K$ inspected $Plot$s and the current $Plot$ associated with the same $Track$, and determines if the time difference between the current $Plot$ and the previous $Plot$ is anomalous. 
An anomalous time difference between $Plot$s indicates malicious $Plot$ dropping or injecting events. 
Such activities may enable an object to evade detection by the detection system or cause a denial of service.

\subsection{\label{subsec:design}Design Considerations}

The following considerations were taken into account in the design of the proposed method: \\
    (1) \textbf{Analysis of heterogeneous data.} As mentioned in Section~\ref{subsec:message_description}, the $Plot$ records consist of numerical and categorical features. 
    Machine learning-based models usually cannot be applied directly to heterogeneous features, so there is a need to transform those features before applying a model.
    
    \noindent (2) \textbf{Immediate detection.}
    Often, radar systems are integrated within real-time decision-making systems. 
    This requires a detection method capable of making a quick and accurate prediction, while keeping the false alarm rate as low as possible. 
    
    \noindent (3) \textbf{Processing sequential (stream) data.}
    As mentioned in Section~\ref{subsec:message_description}, each $Plot$ may be correlated with other $Plot$s (e.g., when they are part of an identified $Track$). 
    Therefore, in addition to capturing and learning patterns among features of the same $Plot$ record, the model should be able to learn patterns within a sequence of $Plot$s. 
    
    \noindent (4) \textbf{Explaining anomalies.} 
    Understanding which part of the data is anomalous can help the system operator perform the correct action. 
    If the method has the ability to identify the packets/features that contribute the most to an anomaly, it can improve the system operators' decision-making capabilities.
    The structure of the proposed method enables it to identify the anomaly type detected, determine whether a legitimate message is missing from a sequence of messages, and understand whether a feature of an existing message has been manipulated.
    

\subsection{Extracted Features} \label{features}
As described in Section \ref{subsec:message_description}, each $Plot$ record contains a variety of features. 
In our dataset, the $Plot$ record includes 10 categorical features and 18 numerical features, all with high potential of being exploited by an attacker interested in disrupting the radar system's normal operation. \\
\noindent \textbf{Categorical features.} Five of the categorical features describe the returned signal's physical properties. 
The other five are: (1) trackType - specifies the $Track$ type, (2) signalQuality - specifies the quality of the returned signal, (3) objectType - specifies the type of object detected, (4) alertRaised - specifies whether an alert has been raised on the system, and (5) objectCategory - specifies whether the object is considered hostile. \\
\noindent \textbf{Numerical features.} One numerical feature is the timestamp indicating when a $Plot$ has been updated on the system. 
The other 17 numerical features ${num1, num2,...,num17}$ relate to the correlations between a detected object's locations.

Note that due to privacy concerns, a more detailed description of the features in the dataset used in this research cannot be provided.

\section{\label{sec:method:details}Proposed Method}

In this section, we provide a detailed description of each component of the proposed method.

\subsection{Field Manipulation Detection}
As illustrated in Figure~\ref{fig:HLA}, this module consists of the following two computational components: (1) the $Plot$-Level Anomaly Detector, and (2) the $Track$-Level Anomaly Detector.
As mentioned earlier, the $Plot$-Level Anomaly Detector generates an anomaly score for each $Plot$ independently, while the $Track$-Level Anomaly Detector aggregates the anomaly scores of the $Plot$s and generates an alert if the value computed exceeds a certain threshold.

\begin{figure}[h]
\centering
\includegraphics[width=0.64\textwidth]{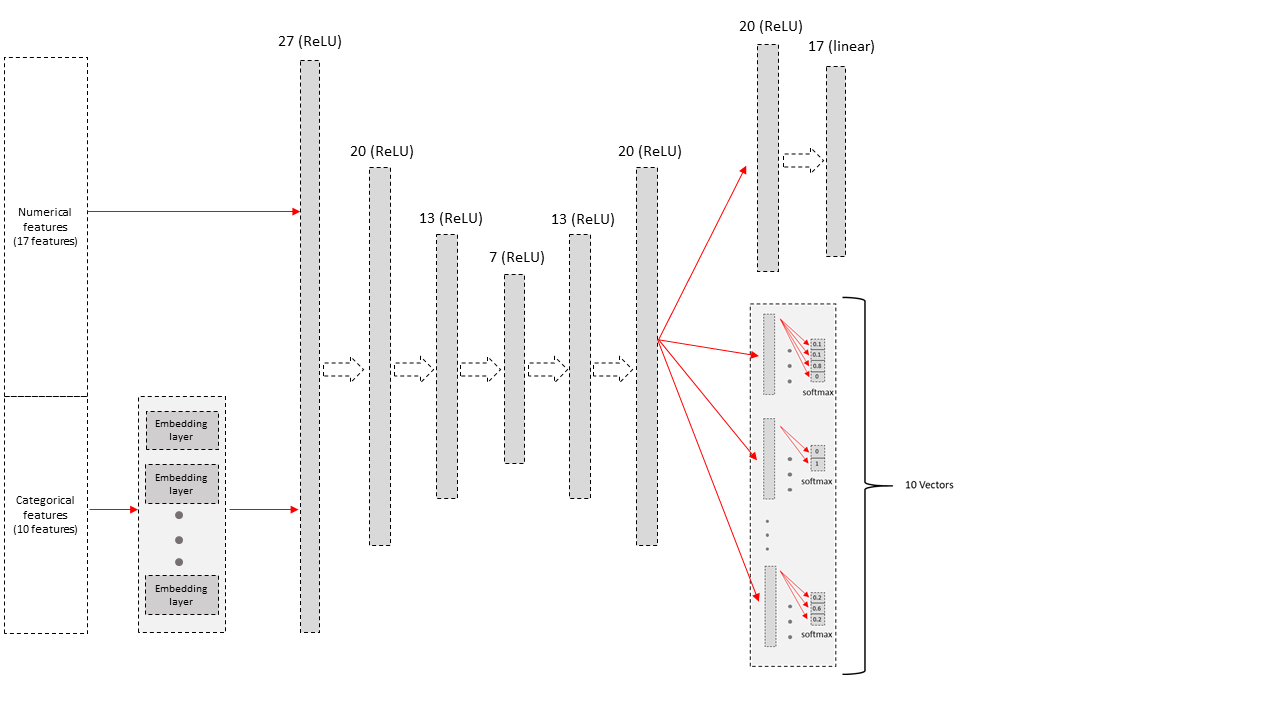}
\caption{The neural network architecture of the proposed \emph{Plot}-Level Anomaly Detector; the network is based on an autoencoder.}
\label{fig:plot_level_anom}
\end{figure}

\textbf{\emph{Plot}-Level Anomaly Detector.}
This module focuses on detecting an anomaly within a single $Plot$. The $Plot$ features used by this module are described in~Section~\ref{features}.
In order to detect an anomaly within a given $Plot$, a special variant of an autoencoder is proposed (Figure~\ref{fig:plot_level_anom}).

The input neurons representing the categorical features are attached with an embedding layer. An embedding technique is commonly used to create a concise, numerical representation of categorical features~\cite{hancock2020survey}. 
The embedding representation of the 10 categorical features is concatenated with the 17 numerical features, resulting in a vector of size 27.
This vector is then fed to the stacked autoencoder consisting of seven layers with respectively 27, 20, 15, 10, 15, 20, and 27 neurons.

Finally, the following output layer is attached: 
\begin{enumerate}
    \item For the numerical features, a fully connected layer is attached, followed by the \emph{linear} activation function. This layer contains 17 neurons against 17 numerical features.
    \item For each categorical feature with $l$ possible values, $l$ output neurons are assigned, followed by the \emph{softmax} function. During inference, this function provides a probability distribution over all possible values for each categorical feature.
\end{enumerate} 

During the training phase, we use data that consists solely of benign $Track$s. As described in Section~\ref{sec:setup}, each $Track$ contains several $Plot$s (training examples). We divide the $Track$s in the dataset so that 80\% is used for training \emph{tr} and 20\% is used for validation \emph{val}. 
As illustrated in Figure~\ref{fig:plot_level_anom}, the label for each example is one vector of 17 entries containing the value of each numerical feature, concatenated to 10 vectors representing the one-hot encoding for each of the 10 categorical features.

The loss function used is the following combination of the \emph{mean squared error} (MSE) and the \emph{sparse categorical cross-entropy} (SCCE):

\begin{equation}
	loss = MSE(x_i: x_i \in numerical) +  \sum_{x_i \in categorical} SCCE(x_i)
\end{equation}

Given \emph{tr}, the network is trained to minimize the loss function on \emph{val} using the Adam optimizer. Training is stopped when the loss function reaches its minimum on \emph{val}.

\textbf{\emph{Track}-Level Anomaly Detector.}
This module consists of a $Track$ Score Collector and a $Track$-Level Anomaly Detector. 
The scores generated by the $Plot$-Level detection module serve as the input for this model. 
The model collects all of the scores for the $Plot$s that are part of the same $Track$ and outputs them for the $Track$-Level Anomaly Detector model. 
All of the $Plot$s' scores for a $Track$ serve as the input to this model.

For each $Track$, we define the anomaly score as the average of all of the input scores calculated by the \emph{Plot}-Level Anomaly Detector (those are stored inside the $Track$ Score Collector). 
In order to determine the anomaly threshold, we used the validation set \emph{val} that was used to train the \emph{Plot}-Level Anomaly Detector. The anomaly threshold is calculated as the maximum value over the $Track$s' average scores.

\subsection{Timing-Based Anomaly Detection}

\begin{figure}[h]
\centering
\includegraphics[width=0.4\textwidth]{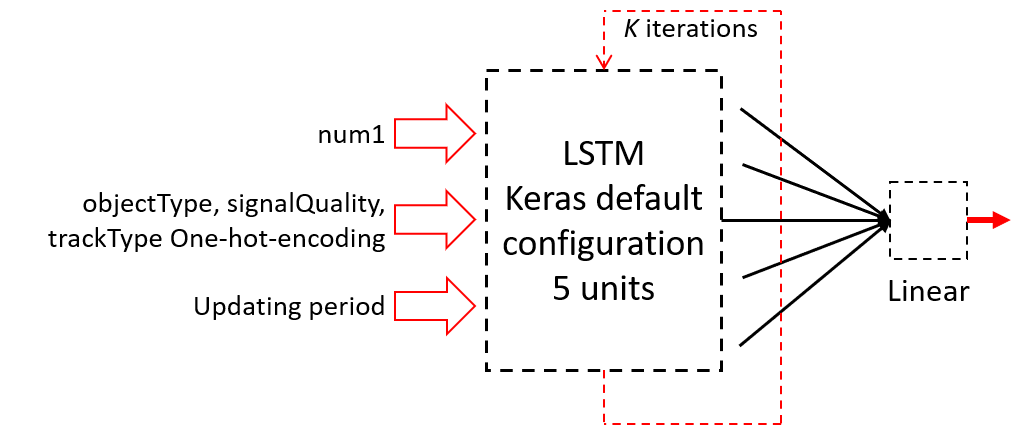}
\caption{The neural network architecture of the proposed Sequence-Based Anomaly Detector. 
The network is based on an LSTM network.}
\label{fig:lstm}
\end{figure}

This module focuses on detecting anomalies related to the $Plot$s' arrival times. It uses the following subset of the $Plot$ features described in Section~\ref{features}: $num1$, objectType, signalQuality, and trackType.
The module consists of the following two components: (1) the Data Preprocessing component, and (2) the Sequence-Based Anomaly Detector (see Figure~\ref{fig:lstm}), which is based on a Long Short-Term Memory (LSTM) network.

\textbf{Data Preprocessing.} 
The input for this component is the current $Plot$, along with the $K$ previous $Plots$ that correspond to the same $Track$ (with a sequence length of $K+1$).
This component consists of the following three parts: 
\begin{enumerate}
    \item Categorical feature one-hot encoding - 
    in this part, each feature is converted to a vector that has a value of one in one entry (corresponding to the feature value) and zeros in all of the other places.  
    \item UpdatingPeriod extraction - in this part, we calculate a feature called UpdatingPeriod. 
    The value of this feature is the time difference between two consecutive $Plot$s in the sequence (e.g., current time - previous time). 
    \item Feature scaling - in this part, we apply min-max scaling on the data for feature scaling. 
\end{enumerate}

\textbf{Sequence-Based Anomaly Detector.}
This module receives a sequence of length $K+1$ as input, and by using the first $K$ elements of the sequence, the model tries to predict the value of the $UpdatingPeriod$  (e.g., time interval) feature of the $K+1$-th element (i.e., current $Plot$). 
To enable this, we designed an LSTM-based regressor.
The architecture of the regressor is presented in Figure~\ref{fig:lstm}. As can be seen, the $K$ length sequence serves as input to an LSTM network with five hidden units. 
Then, the output of the LSTM component in the final step is fully connected to one neuron. 
At the end, a linear activation function is applied to this neuron. 

During the training phase, we use data consisting solely of benign $Track$s. For each $Track$, we generate a set of training examples such that each consists of a set of $K$ consecutive $Plot$s and is labeled by the $UpdatingPeriod$ feature of the following $K+1$-th $Plot$.
We divide the $Track$s in the dataset so that 80\% is used for training \emph{tr} and 20\% is used for validation \emph{val}. The loss function used is the MSE. Given \emph{tr}, the network is trained to minimize the loss function on \emph{val} using the Adam optimizer. Training is stopped when the loss function reaches its minimum on \emph{val}.

Once the model training is complete, the anomaly threshold $thr$ is set. 
This anomaly threshold, above which an instance is considered anomalous, is calculated as the sum of the sample mean and standard deviation of MSE over \emph{val}:

\begin{equation}
	thr = mean(MSE_{val}) + std(MSE_{val}) 
\end{equation}

Regarding the $K$ parameter, the detection accuracy increases as long as $K$ is large. However, an anomaly cannot be detected during the first $K$ $Plot$s. Practically, for radar systems, setting $K$ at five is a good balance in this trade off.
\section{\label{sec:evaluation}Evaluation}

For the evaluation we use benign data collected from four real radar systems that are deployed in different operational setups and used to identify objects. 
We refer to each dataset as a \emph{recording session}.
The number of messages (i.e., radar data records) in each recording session is as follows: R1 -- 45,720 (50.7\%), R2 -- 23,017 (25.5\%), R3 -- 11,955 (13.2\%), and R4 -- 9,551 (10.6\%). 
In total, there are 90,243 messages.

\subsection{\label{subsec:attacks} Attack Scenarios Implemented}

Since our dataset does not include any attacks, we had to simulate the attacks and inject them into our dataset.
We simulated four attack scenarios by manipulating valuable benign data, and conducted 15 different experiments for each attack.

\textbf{Categorical feature manipulation.} 
This is an integrity violation attack in which the attacker changes a categorical feature. 
In the following cases, such manipulations may cause the consuming systems or radar operators to incorrectly characterize and handle the detected object.
\begin{enumerate}
    \item objectType - changing this feature may cause aerial objects to appear as ground objects and vice versa. 
    This can change the way the radar operator reacts to the various objects detected; an attacker can utilize this attack in order to disguise threatening objects, leaving them untreated by radar operators.
    \item alertRaised - changing this feature may cause an alert to be issued for no reason and vice versa.
    Such an attack can create many false alarms, causing the radar operator to ignore threats.
    \item objectCategory - changing this feature may cause friendly objects to be considered enemy objects and vice versa.
    This attack can cause radar operators to fire on friendly objects or ignore enemy objects.
\end{enumerate}

Changing the features for an entire $Track$ or a $Track$ segment will be much more beneficial to the attacker than changing the features for random $Plot$s. 
This is because changes in the features of random $Plot$s may be ignored by the radar operator and are not likely to influence the action taken by an operator. Therefore, we applied the above manipulations on an entire $Track$.

The following process was used to generate the test set given a benign set $B$ collected from real radar systems: (1) create $B'$ by duplicating $B$; (2) select $f$ -- the feature that should be manipulated; (3) for each $Track$ $T$ in $B'$, change the value of $f$ to a distinct random value from the set of valid values of $f$;  (4) label the manipulated $Plot$s as anomalies; and (5) combine the benign set $B$ and the manipulated set $B'$ into one test set.  
This process generates a similar number of benign and malicious $Plot$s; therefore, the resulting test set is balanced.

\textbf{\emph{Plot} dropping.}
This is an availability violation attack in which the attacker drops one/several $Plot$s from a $Track$, so that the detected object will evade the consuming system for some period of time.

Dropping several consecutive $Plot$s will be much more beneficial to the attacker than dropping nonconsecutive $Plot$s or just a single $Plot$. Therefore, for a $Track$ $T$, we first select a minimum value $c$ ($c\le|T|$) of $Plot$s to be removed from $T$. We use $c=10$ since a value less than 10 does not allow an attacker to cause real harm to the data integrity of a radar system. 

The following process was used to generate the test set given the benign set $B$ collected from real radar systems: (1) for each $Track$ $T$, select a random $Plot$ index $i$ and a random integer $r\in{c, |T|-1}$; (2) starting at index $i$, drop a minimum number $minimum(|T|-i, r)$ of consecutive $Plot$s; (3) label the $Plot$ that follows the dropped $Plot$s as an anomaly.
This process generates a different number of benign and malicious $Plot$s; therefore the resulting test set is imbalanced.

\subsection{\label{subsec:evalsetup}Evaluation Method}

We conducted three types of evaluations; in each case, we repeated the experiment five times (each time a different recording session was selected for testing).

A description of the types of evaluations performed is provided below, along with an example of the training/testing data in a scenario in which the R4 recording session is examined: \\
\begin{itemize}
    \item \textbf{Cross-session setup:} in this case, each time we used three recording sessions for training, and the fourth session was used to test the model. 
    Example for the R4 recording session: training -  R1, R2, and R3; testing - R4. 
    \item \textbf{Chronological setup:} in this case, for each recording session, we trained the model on the first 90\% of the instances (in chronological order) and tested it on the remaining 10\% of the instances.
    Example for the R4 recording session: training - 90\% of the instances of R4; testing - remaining 10\% of the instances of R4.
    \item \textbf{Transfer learning setup:} this case is a combination of the previous two setups, in which we used three recording sessions, as well as the first 10\% of the instances of the fourth session, to train the model, and the remaining 90\% of the instances of the fifth session were used to test the model. 
    Example for the R4 recording session: training -  R1, R2, R3, and the first 10\% of the instances of R4; testing - the remaining 90\% of the instances of R4.
\end{itemize}
It should be noted that the cross-session evaluation setup has two significant advantages over the other two: (1) in practice, it is easier for radar engineers to deploy a model that has already been trained rather than training a model using new system data, and (2) training the model on new system data can expose the model to cyber risks (e.g., adversarial poisoning~\cite{vorobeychik2018adversarial}).

\subsection{\label{subsec:evalmetric}Evaluation Metrics}

The task of identifying anomalies in the radar data stream is a binary classification task, where benign samples are labeled as zero (negative), and malicious data is labeled as one (positive).
We used the following common metrics for the evaluation: true positive rate (TPR)/recall, false positive rate (FPR), receiver operating characteristic (ROC) curve, area under the ROC curve (AUC), precision, precision-recall curve, and average precision (AP).

\subsection{\label{subsec:results}Results}
In this section, we present the evaluation results for each attack scenario (presented in Section~\ref{subsec:attacks}) and evaluation method (presented in Section~\ref{subsec:evalsetup}).
First, we present a graph of the evaluation results as a function of different thresholds (using ROC and PRC graphs). 
Then, we summarize the evaluation results for predefined anomaly thresholds. 
The method for calculating the anomaly thresholds is described in Section~\ref{sec:method:details}. 

\begin{figure*}[h]
\centering
\includegraphics[width=0.55\textwidth]{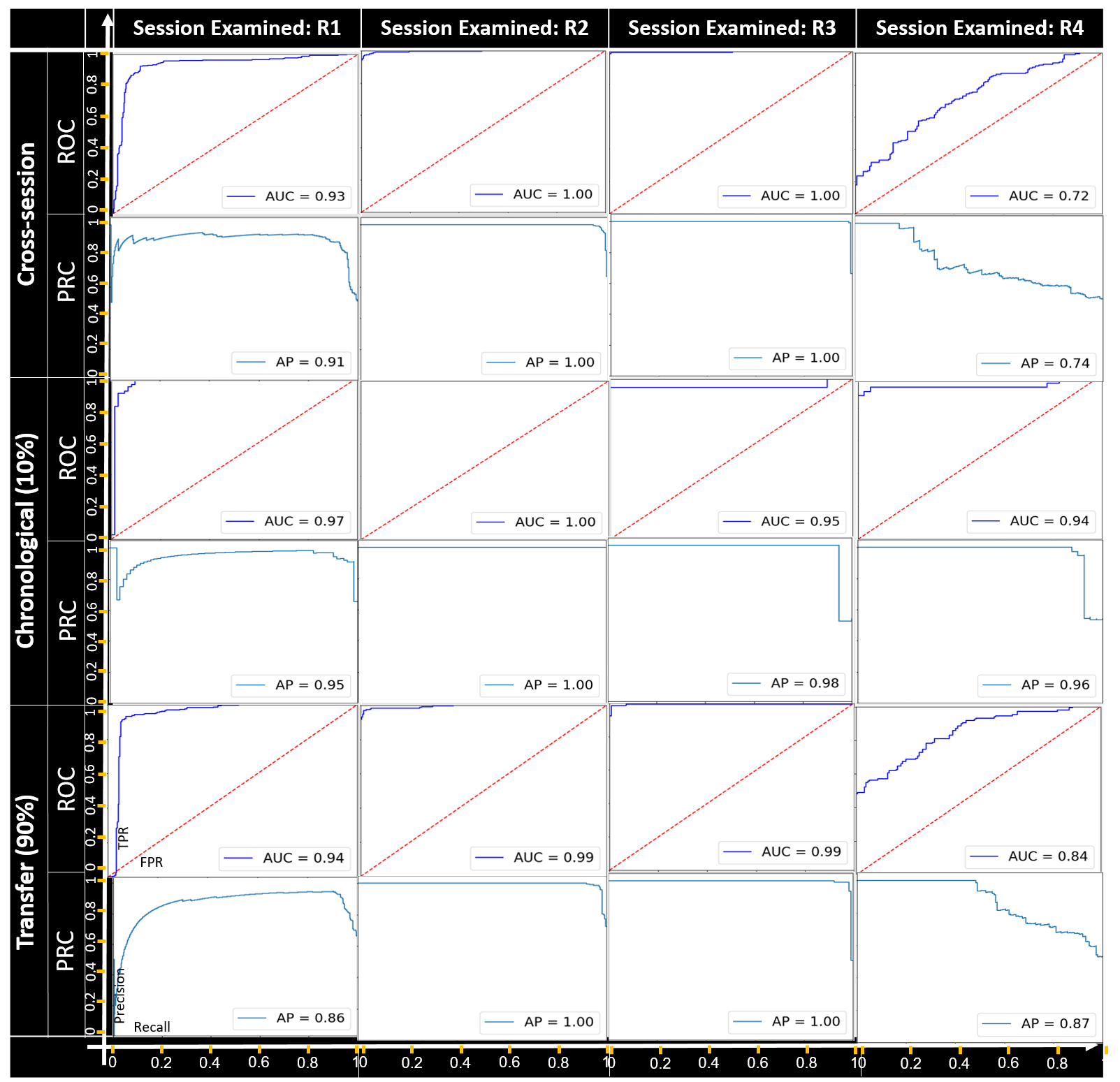}
\caption{The ROC and PRC curves for the objectType feature manipulation attack for each recording session examined.}
\label{fig:objectType}
\end{figure*}

\begin{figure*}[h]
\centering
\includegraphics[width=0.55\textwidth]{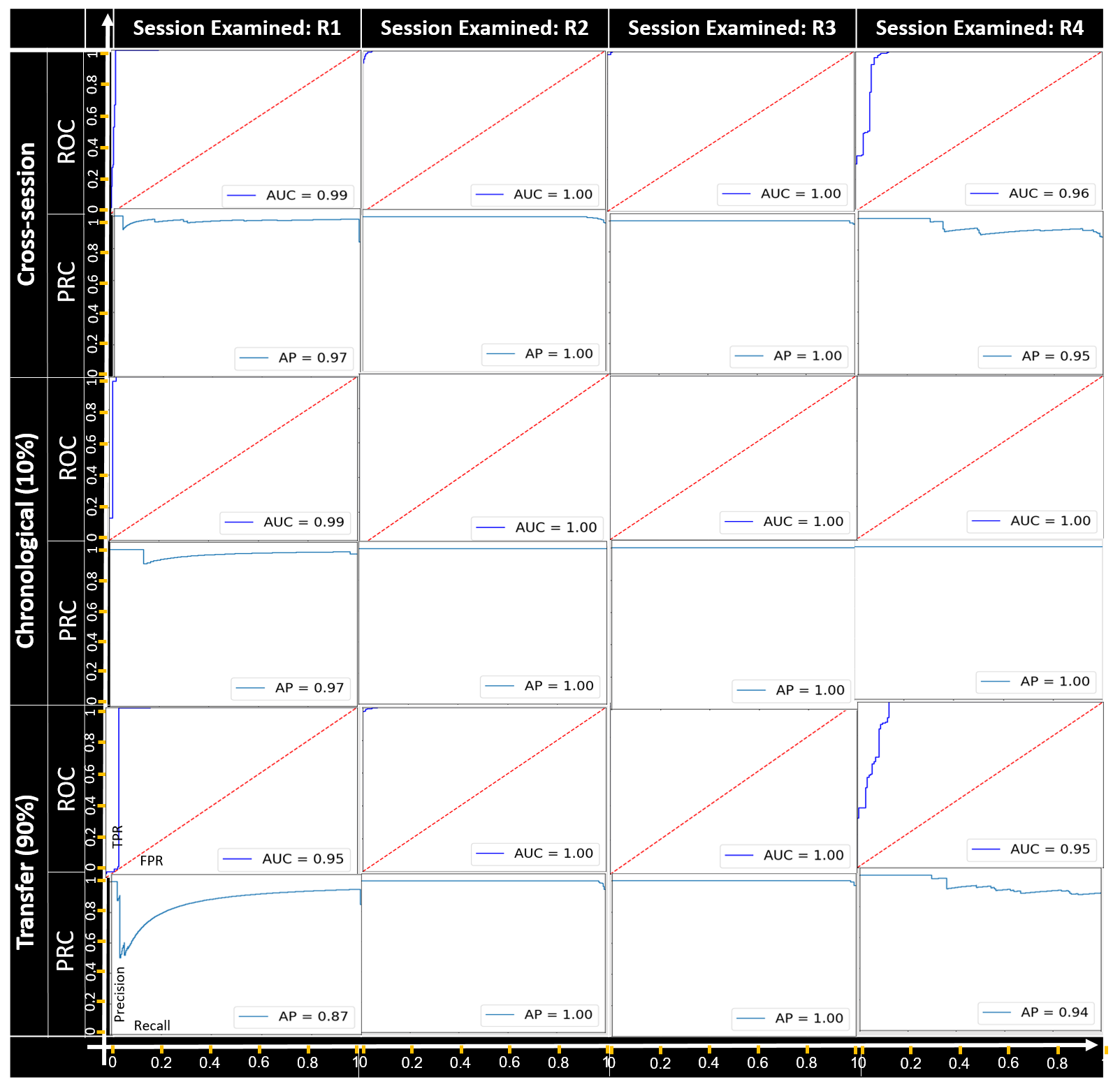}
\caption{The ROC and PRC curves for the objectCategory feature manipulation attack for each recording session examined.}
\label{fig:objectCategory}
\end{figure*}

\begin{figure*}[h]
\centering
\includegraphics[width=0.55\textwidth]{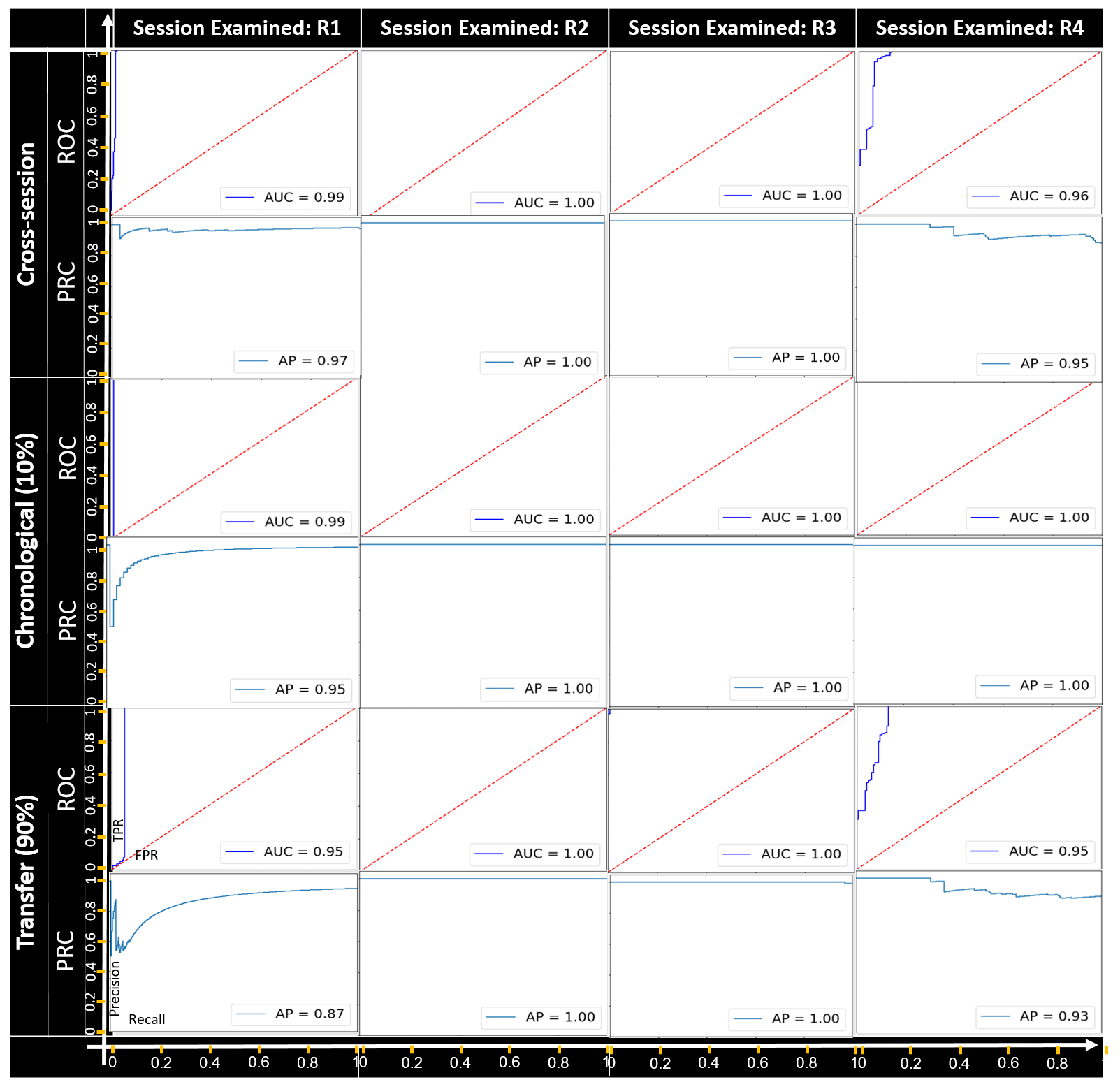}
\caption{The ROC and PRC curves for the alertRaised feature manipulation attack for each recording session examined.}
\label{fig:alertRaised}
\end{figure*}

\noindent \textbf{Categorical feature manipulation.}
Figures~\ref{fig:objectType} -~\ref{fig:alertRaised} present the detection results of the Field Manipulation Detection module when manipulating objectType, objectCategory, and alertRaised features respectively.
As can be seen, in all cases except for recording session R4 of the objectType feature manipulation attack, all setups had great performance in terms of the AUC and AP for all categorical feature manipulation attacks.
For recording session R4 of the objectType feature, the best performance was obtained with the chronological setup.
The best detection rates were observed for the manipulation attack on both objectCategory and alertRaised features. 

\begin{figure*}[h]
\centering
\includegraphics[width=0.55\textwidth]{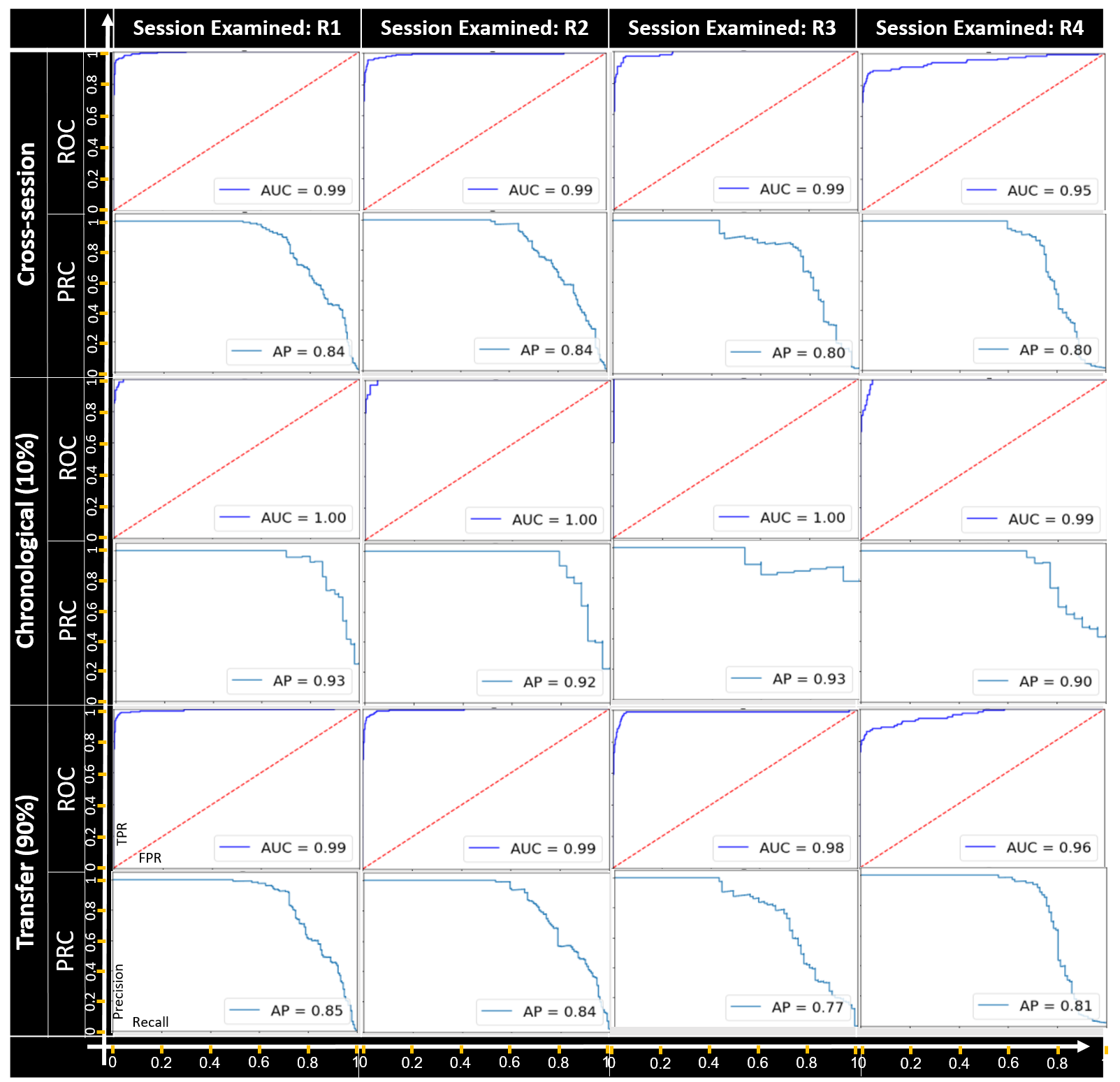}
\caption{The ROC and PRC curves for the \emph{Plot} dropping attack for each recording session examined.}
\label{fig:drop_packet}
\end{figure*}

\noindent\textbf{\emph{Plot} dropping.}
Table~\ref{tab:drop_packet} presents the distribution of the benign /

\noindent malicious samples in the $Plot$ dropping attack. 
It is noteworthy that the data is imbalanced, since for each $Track$ we only execute the attack once. 
The performance of the Timing-Based Anomaly Detection module on the generated testset is presented in Figure~\ref{fig:drop_packet}.
As can be seen, in all scenarios the model achieved good results. 

\noindent\begin{minipage}{\linewidth}
\noindent\textbf{Performance for predefined anomaly thresholds.}
Table~\ref{tab:results_summary} presents the performance for each attack scenario and each setup in terms of the true positive rate (TPR) and false positive rate (FPR) for a threshold that was computed on a validation set and then applied on a test set.
As can be seen, our proposed method can detect 90\% of the $Plot$ dropping attacks, with a false positive rate of 2\%, and from 76\% to 96\% of feature manipulation attacks, with a false positive rate of 2\%. 
In typical scenarios, 400 $Plot$s are generated and transferred by a radar system every minute. A false alarm rate of 2\% means that eight $Plot$s should be tracked by a system operator/automatic system every minute, which is practical.
\end{minipage}

In addition, in most cases our proposed method achieves very good results for the cross-session setup. This is an important point, since it means that a pretrained model can be migrated to new radar systems without retraining.

\begin{table}[h]
\centering
\caption{Number of benign (b) and malicious (m) samples in the \emph{Plot} dropping attack.}
\scriptsize
\begin{tabular}{|l|c|c|c|c|c|c|c|c|}
\hline
\multirow{3}{*}{} & \multicolumn{8}{c|}{Recording Session Examined} \\ 
\cline{2-9}  & \multicolumn{2}{c|}{R1}  & \multicolumn{2}{c|}{R2}  & \multicolumn{2}{c|}{R3}  & \multicolumn{2}{c|}{R4} \\ 
\cline{2-9}  & b & m & b & m & b & m & b & m \\ \hline
Cross-session  & 46,600 & 213 & 23,068 & 177 & 12,139  & 97 & 9,484 & 119 \\ \hline
Chron. (10\%) & 4,606 & 60 & 2,323 & 34  & 1,206 & 15 & 842 & 31 \\ \hline
Transfer (90\%) & 42,709 & 195 & 21,126 & 170 & 10,808 & 86 & 8,451 & 110 \\ \hline
\end{tabular}
\label{tab:drop_packet}
\end{table}

\begin{table}[h]
\caption{Evaluation results obtained for predefined anomaly thresholds.}
\centering
\scriptsize
\begin{tabular}{|l|r|r|r|r|}

\hline
\textbf{Setup} & \multicolumn{1}{l|}{\textbf{Avg AUC}} & \multicolumn{1}{l|}{\textbf{Avg PR}} & \multicolumn{1}{l|}{\textbf{Avg TPR}} & \multicolumn{1}{l|}{\textbf{Avg FPR}} \\ \hline
\multicolumn{5}{c}{Feature manipulation - objectType} \\ \hline
Cross-session    & 0.922                              & 0.930                              & 0.759                              & 0.028                             \\ \hline
Chronological (10\%) & 0.972                              & 0.712                             & 0.894                              & 0.016                             \\ \hline
Transfer (90\%)      & 0.943                              & 0.931                             & 0.710                              & 0.090                             \\ \hline
\multicolumn{5}{c}{Feature manipulation - objectCategory} \\ \hline
Cross directories    & 0.986                              & 0.973                              & 0.906                              & 0.028                             \\ \hline
Chronological (10\%) & 0.997                              & 0.994                             & 0.770                              & 0.016                             \\ \hline
Transfer (90\%)      & 0.975                              & 0.952                             & 0.769                              & 0.090                             \\ \hline
\multicolumn{5}{c}{Feature manipulation - alertRaised} \\ \hline
Cross-session    & 0.987                              & 0.975                             & 0.963                              & 0.028                             \\ \hline
Chronological (10\%) & 0.997                              & 0.988                             & 0.776                              & 0.016                             \\ \hline
Transfer (90\%)      & 0.974                              & 0.950                              & 0.750                                  & 0.090                             \\ \hline
\multicolumn{5}{c}{\emph{Plot} dropping} \\ \hline
Cross-session    & 0.975                              & 0.820                             & 0.900                              & 0.023                             \\ \hline
Chronological (10\%) & 0.987                              & 0.916                             & 0.937                              & 0.026                             \\ \hline
Transfer (90\%)      & 0.983                              & 0.830                             & 0.909                              & 0.023                             \\ \hline
\end{tabular}
\label{tab:results_summary}
\end{table}

\subsection{\label{subsec:preformance}Performance Analysis}

In order to understand whether the proposed method is practical and can be applied for real-time tracking of anomalies in radar system data streams, we measured the average time it takes to process a single $Plot$.
This experiment was conducted on a standard machine with an Intel core i7-7600U 2.8Ghz CPU and 32GB RAM.
The machine's operating system was Windows 10. 

The results showed that our proposed method can analyze 20K $Plot$s per second.
Given that in typical scenarios a radar system generates and transfers 400 $Plot$s per minute, we conclude that our method is practical and can be used for real-time tracking of anomalies in radar system data streams.
\section{\label{sec:summary}Summary \& Conclusions}

In this paper, we propose a novel deep learning-based method for detecting data manipulation attacks on radar systems. 
We consider attack scenarios in which an attacker manipulates the data sent by the radar system to peripheral consuming systems.
To evaluate our method, we used data collected from four real radar systems. 
This dataset consists of $Plot$ records that are sent from the radar system to a peripheral consuming system.  

To evaluate our method, we generated attacks considered beneficial to the attacker, ranging from integrity violation attacks to availability violation attacks.
The results of our evaluation show that the proposed method can learn the data's normal behavior and distinguish between normal and manipulated data.

In our design process, we took into account the benefits of identifying the anomaly type detected, determining whether a legitimate $Plot$ is missing from a sequence of $Plot$s, and understanding whether 
an existing $Plot$'s feature has been manipulated.
This property could help radar operators determine the correct action to take when an attack occurs.

The transferability and practicality of the proposed method is demonstrated by: (1) our method obtains very good results when training on data collected from three radar systems and testing on data collected from another radar system, and (2) a typical Intel controller can analyze $Plot$s at a frequency of 20K/sec, while radar systems generate and transfer $Plot$s at a frequency of 7/sec.

\FloatBarrier
\bibliographystyle{ACM-Reference-Format}
\bibliography{ref}


\end{document}